\documentclass[doublecol]{epl2}

\usepackage{ifthen}
\usepackage{ifpdf}
\usepackage{color}

\ifpdf
\usepackage{graphicx}
\usepackage{epstopdf}
\else
\usepackage{graphicx}
\usepackage{epsfig}
\fi

\usepackage{latexsym}
\usepackage{amsmath}
\usepackage{amssymb}
\usepackage{bm}
\usepackage{wasysym}

\newcommand{\barline}[1]{#1}

\newcommand{\eexp}{\mbox{e}^}

\newcommand{\amatrix}[1]{\begin{matrix} #1 \end{matrix}}

\newcommand{\mylabel}[1]{\label{#1}} 
\newcommand{\beq}{\begin{eqnarray}}
\newcommand{\eeq}{\end{eqnarray}} 
\newcommand{\be}[1]{\begin{eqnarray}\ifthenelse{#1=-1}{\nonumber}{\ifthenelse{#1=0}{}{\mylabel{e#1}}}}
\newcommand{\ee}{\end{eqnarray}} 

\newcommand{\Eq}[1]{\textcolor{blue}{Eq.\!\!~(\ref{#1})}} 
\newcommand{\Fig}[1]{\textcolor{blue}{Fig.}\!\!~\ref{#1}} 
\newcommand{\hide}[1]{} %{\textcolor{red}{[hidden text]}} %{}

\newcommand{\widebar}[1]{\overline{#1}}

\title{The non-equilibrium steady state of sparse systems \\ with nontrivial topology}
\shorttitle{NESS of sparse systems}

\author{Daniel Hurowitz$^1$, Saar Rahav$^2$, and Doron Cohen$^1$}

\institute{
\mbox{$^1$Department of Physics, Ben-Gurion University of the Negev, Beer-Sheva 84105, Israel}\\
\mbox{$^2$Schulich Faculty of Chemistry, Technion - Israel Institute of Technology, Haifa 32000, Israel}
}

\abstract{
We study the steady state of a multiply-connected system 
that is driven out of equilibrium by a sparse perturbation.
The prototype example is an $N$-site ring coupled to a thermal bath, 
driven by a stationary source that induces transitions with log-wide distributed rates.
An induced current arises, which is controlled by the strength of the driving, 
and an associated topological term appears in the expression for the energy absorption rate. 
Due to the sparsity, the crossover from linear response to saturation 
is mediated by an intermediate regime, where the current is exponentially 
small in~$\sqrt{N}$, which is related to the work of Sinai on ``random walk in a random environment".  
}

\pacs{}{}

\begin{document}
\maketitle

%%%%%%%%%%%%%%%%%%%%%%%%%%%%%%%%%%%%%%%%%%%%%%%%%%%%%%%%%%%%%%%%%%%%%%%%%%%%%%%%%%%%%%%%%%
%%%%%%%%%%%%%%%%%%%%%%%%%%%%%%%%%%%%%%%%%%%%%%%%%%%%%%%%%%%%%%%%%%%%%%%%%%%%%%%%%%%%%%%%%%

The transport in a chain due to non-symmetric transition 
probabilities is a fundamental problem in statistical 
mechanics \cite{derrida,sinai,d1,d2,d3,d4}. 
It can be regarded as {\em a random walk in a random environment}. 
The seminal observation is due to Sinai \cite{sinai}: 
considering a chain of length $N$, the randomness implies 
a buildup of an exponentially large potential barrier $\exp(\sqrt{N})$, 
and consequently an exponential suppression of the current, 
reflecting a sub-diffusive $[\log(t)]^4$ spreading in time.   

We would like to explore how this picture is modified
if the chain is replaced by a configuration with a non-trivial topology, 
such as a ring, accounting for: 
(1)~unavoidable telescopic {\em correlations} between the transition probabilities;  
(2)~{\em sparsity} due to log-wide distribution of the transition rates as in glassy systems.

\begin{figure}

\centering
\includegraphics[width=0.6\hsize]{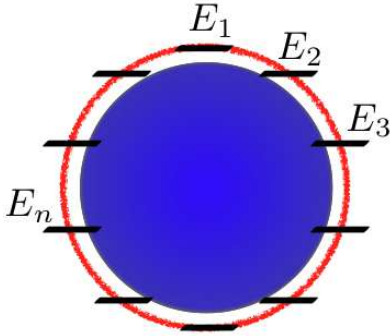}

\caption{
A ring made up of $N$ isolated sites with on site energies $\epsilon_n$.
The ring is coupled to a heat reservoir (represented by the blue "environment") and subjected
to a noisy driving field (represented by the red circle) that induces a current in the ring.
In the numerical tests the energies occupy a band of width ${\Delta=1}$, 
and the temperature of the bath is ${T_{B}=2}$.
The driving source induced rates $w_{\widebar{n}}^{\epsilon}$
are log-box distributed over 8~decades, 
while the bath induced rates are all with ${w_{\widebar{n}}^{\beta}=1}$. 
} 

\label{f0}
\end{figure}

Let us expand on the roles played by these two ingredients:  
In the physical model that we would like to consider (\Fig{f0})
the transition rates are not independent random variables. 
Rather they are determined by {\em differences} of uncorrelated on-site energies. 
These differences are strongly correlated: their sum does 
not grow with~$N$. Consequently the theory of Sinai does not apply.
We shall see that in order to witness a ``Sinai regime" we have 
to introduce into the model an additional ingredient: 
one may call it {\em glassiness} or {\em sparsity}. 
These terms emphasize complementary characteristics 
of the assumed log-wide distribution of the transition rates: 
the distribution span several decades; accordingly the vanishingly 
small values constitute the majority; while the large values 
constitute a minority; which is reflected by having a median 
that is much smaller compared with the algebraic average.

\begin{figure}[h!]

\includegraphics[width=\hsize,clip]{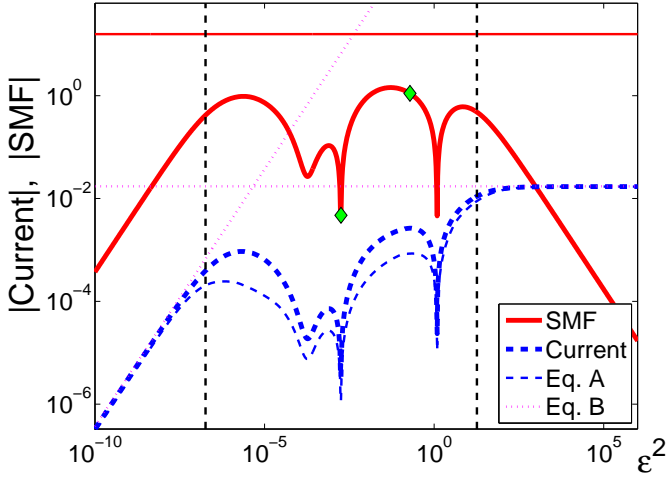}

\caption{We consider the Ring of \Fig{f0} with $N{=}10^{3}$ sites.
The absolute values of the SMF (solid curve) and the current (thick dashed curve) 
are plotted as a function of the scaled driving intensity.   
The vertical dashed lines $1/g_{\max}$ and $1/g_{\min}$ bound 
the Sinai regime. Due to the sparsity ($g_{\min} \ll g_{\max}$) 
it extends over several decades. 
The solid horizontal line is the estimates for the SMF \Eq{e131}.
The two straight dotted lines (labeled as Eq.B) are the estimates 
for~$I$ based on \Eq{e55}, while the dashed line (labeled as Eq.A) 
is the global approximation \Eq{e54}.
The diamonds indicate valued for which we plot the potential  
landscape in \Fig{f1b}. 
}

\label{f1a}
\end{figure}

Our focus is on the non-equilibrium steady state (NESS) 
global current~$I$ that circulates the whole ring,  
and on the associated energy absorption rate (EAR). 
Let us present some numerical results that clarify the physical 
picture and motivate the subsequent analysis. We consider 
a ring that is composed of~$N$ sites (\Fig{f0}). 
The ring is weakly coupled to a bath that has temperature~$T_B$. 
In the absence of driving the average current~$I$ is zero. 
The driving is modeled as a stationary noise source that has an intensity $\epsilon^2$. 
The driving breaks detailed balance, leading to a non-vanishing
affinity along the ring. This affinity, which we call {\em stochastic motive force} (SMF), 
labeled as $\mathcal{E}_{\circlearrowleft}$,  
induces a circulating steady-state current, see \Fig{f1a}.  
For weak driving one observes a linear response behavior 
${I \propto  \mathcal{E}_{\circlearrowleft} \propto \epsilon^2}$.
For very strong driving $I$ saturates.

Both the linear response for weak driving and 
the saturation for strong driving have a relatively 
simple explanation (see details later). 
But the crossover from linear response to saturation 
is more subtle, because it is related to the distribution 
of the transitions rates.
If the distribution is log-wide we call the system {\em sparse}, 
implying that most of elements are much smaller compared with the average. 
Due to the sparsity one observes that there 
is an intermediate region where $\mathcal{E}_{\circlearrowleft} \sim \sqrt{N}$, 
while the current becomes exponentially small, 
exhibiting fluctuations as a function of~$\epsilon^2$.

%%%%%%%%%%%%%%%%%%%%%%%%%%%%%%%%%%%%%%%%%%%%%%%%%%%%%%%%%%%%%%%%%%%%%%%%%%%%%%%%%%%%%%%%%%
\vspace{5mm}
{\bf Outline.-- } 
We introduce the Ring model and clarify its relation 
to the standard paradigm of NESS analysis. 
We derive expressions for the SMF, for the current, 
and for the EAR, illuminating their dependence 
on the driving intensity. With regard to the EAR,  
we highlight the manifestation of a topological term.

%%%%%%%%%%%%%%%%%%%%%%%%%%%%%%%%%%%%%%

%%%%%%%%%%%%%%%%%%%%%%%%%%%%%%%%%%%%%%%%%%%%%%%%%%%%%%%%%%%%%%%%%%%%%%%%%%%%%%%%%%%%%%%%%%%%%%%
\vspace{5mm}
{\bf Sparse networks.-- }
Consider a general rate equation 
for the occupation probabilities~$p_n$, 
namely
\be{1000}
\frac{dp_n}{dt} \ = \ \sum_m w_{nm}p_m - w_{mn}p_n
\ee
where $w_{nm}$ is the transition rate from~$m$ to~$n$. 
We can regard it as describing a network that consists of "sites" connected by "bonds". 
Specifically we consider later a ring that consists of $N$ sites (\Fig{f0}). 
With regard to the bond ${x \equiv (m \leadsto n)}$, that connects site~$m$ to site~$n$, 
we define the coupling $w(x)$ and the field $\mathcal{E}(x)$ as follows: 
\be{100}
\mathcal{E}(x) &\equiv& \ln \left[\frac{w_{nm}}{w_{mn}}\right],
\\
w(x) &\equiv& \left[ w_{nm}w_{mn} \right]^{1/2} 
\label{eSMF}
\ee
We say that the system is {\em sparse} or {\em glassy} 
if either $w(x)$ or $\mathcal{E}(x)$ of the connecting bonds 
have a log-wide distribution.  
This means that there is a small fraction of bonds 
where the coupling or the field is strong, 
while in the overwhelming majority it is very small.

We can regard $\mathcal{E}(x)$ as the ``potential difference"
across the bond ${x = (m \leadsto n)}$. 
The reason for this terminology becomes obvious 
if one considers bath induced transition [see later discussion regarding \Eq{e1}]:  
Then it equals $-(E_n-E_m)/T_B$. 
Note that optionally $\mathcal{E}(x)$ can be
regarded as the entropy change of the bath during 
the ${(m \leadsto n)}$ transition~\cite{eprd}. 
Consequently it is natural to define the ``potential variation" between   
two points $x_1$ and $x_2$ along a line segment as follows:
\be{102}
\mathcal{E}(x_1 \leadsto x_2) \ \ = \ \ \int_{x_1}^{x_2} \mathcal{E}(x) dx 
\ee
Given a loop, one defines the SMF (or mesoscopic affinity~\cite{net1}) as follows:
\be{101}
\mathcal{E}_{\circlearrowleft} \ \ \equiv \ \  \sum_{x} \mathcal{E}(x)  \ \ \equiv \ \ \oint \mathcal{E}(x) dx 
\ee
The summation above is over the bonds~$x$ along the loop, 
which becomes an integral in the continuum limit. 
For a detailed balanced system the SMF is zero for any loop. 
Otherwise the system relaxes to a NESS. 

\begin{figure}[h!]

\includegraphics[width=\hsize,clip]{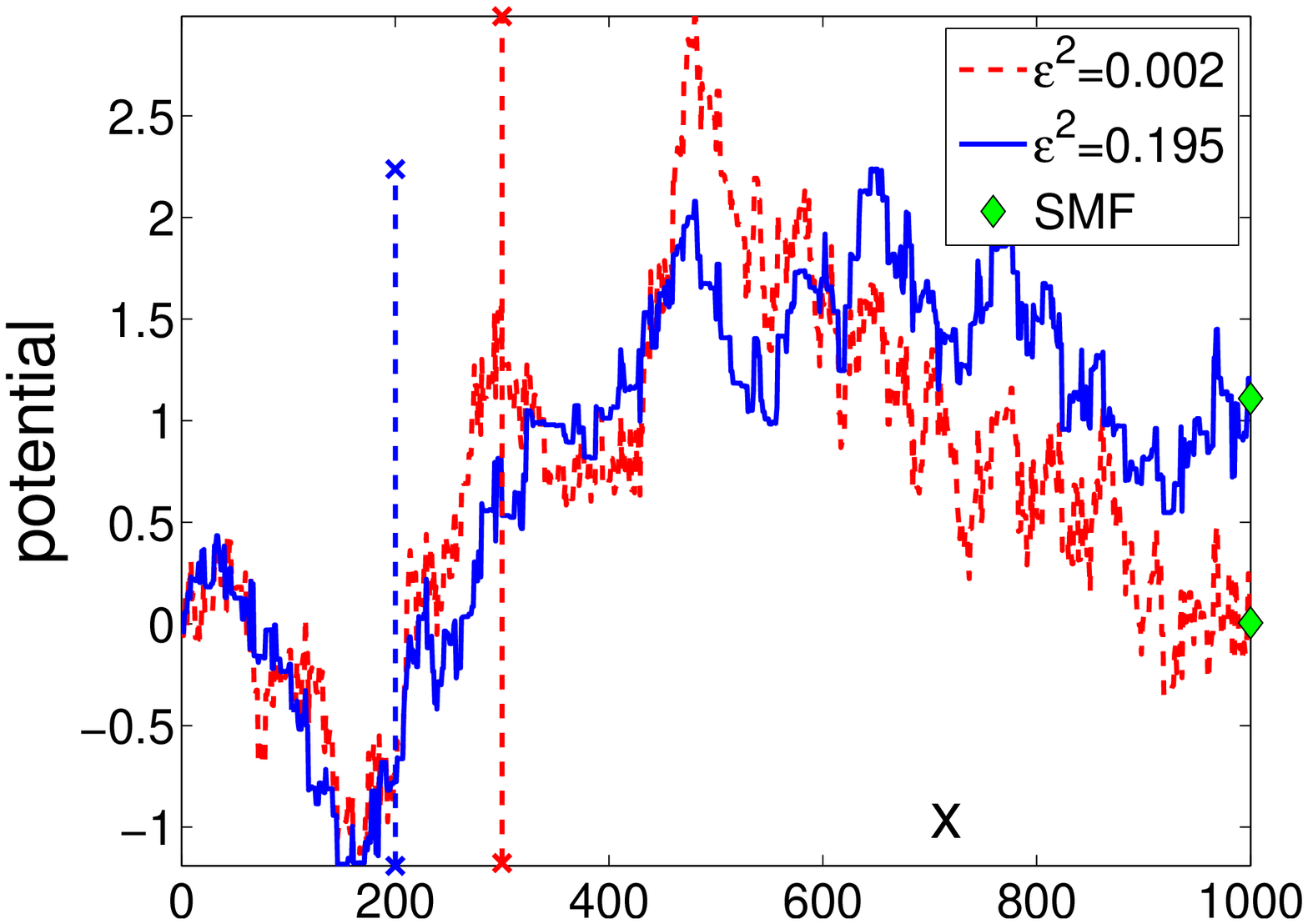}

\caption{The Ring of \Fig{f1a} is considered.  
The potential difference ${\mathcal{E}(0 \leadsto x)}$ 
is plotted against~$x$, for two values of the driving intensity.
The vertical lines correspond to the maximal potential variation $\mathcal{E}_{ \cap}$. 
The SMF values $\mathcal{E}_{ \circlearrowleft}\equiv \mathcal{E}(0 \leadsto N)$ 
are indicated by diamonds.
\hide{
For $\epsilon^{2}=0.18$ we see that $\mathcal{E}_{ \cap} = 4.17$ and $\mathcal{E}_{ \circlearrowleft} = 0.005$.
For $\epsilon^{2}=23.41$ we see that ${\mathcal{E}_{ \cap} = 3.32}$ and ${\mathcal{E}_{ \circlearrowleft} = 1.018}$.} 
}

\label{f1b}
\end{figure}

For the subsequent analysis we define ``the effective potential barrier" 
along a segment as the maximal potential variation (see how it looks in \Fig{f1b}):
\beq
\mathcal{E}_{\cap} \ \ \equiv \ \  \mbox{maximum}\Big\{ |\mathcal{E}(x_1 \leadsto x_2)| \Big\}  
\eeq 
Referring to a Ring, it is important to realize that $\mathcal{E}_{\cap}$ 
cannot be smaller than  $\mathcal{E}_{\circlearrowleft}$. 
If the $\mathcal{E}(x)$ were totally uncorrelated 
both the maximal potential variation and the SMF would be proportional to $\sqrt{N}$.

%%%%%%%%%%%%%%%%%%%%%%%%%%%%%%%%%%%%%%%%%%%%%%%%%%%%%%%%%%%%%%%%%%%%%%%%%%%%%%%%%%%%%%%%%%
\vspace{5mm}
{\bf NESS paradigm.-- }
In the physical problem the network consists of $N$ sites, 
with on-site energies $E_n$. The transition rates $w_{nm}$ 
from site~$n$ to site~$m$ are induced by a driving source 
that has an intensity~$\epsilon^2$, and by a bath that 
has a temperature~$T_{B}$. Namely,  
\be{1}
w_{nm} \ \ = \ \ w^{\epsilon}_{nm} + w^{\beta}_{nm}
\ee
where ${w^{\epsilon}_{nm} = w^{\epsilon}_{mn}  \propto  \epsilon^2}$, 
while the bath is detailed-balanced with $w^{\beta}_{nm}/w^{\beta}_{mn} = \exp[-(E_n{-}E_m)/T_B]$.
One can define a symmetrized rate 
as ${w_{\overline{nm}} = (w_{nm} + w_{mn})/2}$. 
Then it follows that 
\be{111}
w^{\beta}_{nm} \ \ = \ \ \frac { 2w^{\beta}_{\overline{nm}} } {1 + \exp\left(\frac{E_n-E_m}{T_B}\right)}
\eeq

Our setting as described above is formally a special case 
of the common non-equilibrium paradigm for a system 
that is coupled to two heat baths.
The driving source is like a bath that has temperature ${T_A=\infty}$, 
while the environment is a bath that has a finite temperature ${T_B<\infty}$.
As discussed in Ref.\cite{derrida} the generalization of the detailed 
balance condition requires 
\be{0}
\frac{w_{nm}(Q_A,Q_B)}{w_{mn}(-Q_A,-Q_B)}  \ \ = \ \ \exp\left[-\frac{Q_A}{T_A} -\frac{Q_B}{T_B} \right]
\ee
Where $Q_A$ and $Q_B$ are the the amounts of energies that are 
transferred from the baths to the system. In our setting 
the two baths are independent of each other and therefore 
any event is either with ${Q_A=E_n-E_m}$  and ${Q_B=0}$, 
or with ${Q_B=E_n-E_m}$  and ${Q_A=0}$.

%%%%%%%%%%%%%%%%%%%%%%%%%%%%%%%%%%%%%%%%%%%%%%%%%%%%%%%%%%%%%%%%%%%%%%%%%%%%%%%%%%%%%%%%%%
%%%%%%%%%%%%%%%%%%%%%%%%%%%%%%%%%%%%%%%%%%%%%%%%%%%%%%%%%%%%%%%%%%%%%%%%%%%%%%%%%%%%%%%%%%
\vspace{5mm}
{\bf Microscopic temperature.-- }
In the absence of driving the transitions that are induced 
by the bath satisfy detailed balance, and accordingly  
the steady state of the system is {\em canonical} 
with occupation probabilities ${p_n \propto \exp[-E_n/T_B]}$. 
Once we add the driving this is no longer true.
Still, there is a well defined NESS, 
so we can formally define a microscopic temperature 
for each transition separately via the formula
\be{71}
\frac{p_n}{p_m} \ \ = \ \ \exp\left[-\frac{E_n-E_m}{T_{nm}} \right]
\eeq
Unlike a canonical state, here we may have 
a wide distribution of microscopic temperatures.
Furthermore, in a NESS the local temperature 
can be negative, i.e. the occupation of a higher level 
can be larger than the occupation of a lower one.

%%%%%%%%%%%%%%%%%%%%%%%%%%%%%%%%%%%%%%%%%%%%%%%%%%%%%%%%%%%%%%%%%%%%%%%%%%%%%%%%%%%%%%%%%%
\vspace{5mm}
{\bf The Ring model.-- }
As a specific example we consider a ring 
with random on-site energies $E_n \in [0,\Delta]$,  
and near neighbor transitions. 
We use the notations $\Delta_n=E_{n}-E_{n{-}1}$, 
and ${w_{\overrightarrow{n}}=w_{n,n{-}1}}$, and ${w_{\overleftarrow{n}}=w_{n{-}1,n}}$, 
and ${w_{\widebar{n}}=(w_{\overrightarrow{n}}+w_{\overleftarrow{n}})/2}$. 
The superscripts $\beta$ and $\epsilon$ are used 
in order to distinguish the bath and driving source contributions. 
It is useful to notice that in the high temperature limit \Eq{e111} 
takes the form
\beq
\left\{ w^{\beta}_{\overrightarrow{n}}, \,  w^{\beta}_{\overleftarrow{n}}  \right\}
\ \ = \ \ 
\left[ 1 \mp \frac{\Delta_n}{2T_B} \right] w^{\beta}_{\widebar{n}} 
\eeq

The rate equation \Eq{e1000} is formally identical 
to an electrical network where the $p_n$ are like voltages, 
and the $w_{nm}$ represent conductors.
Inspired by this formal analogy, we regard the conductivity  
of the network as an ``average" transition rate, which we call $\barline{w}$.
Consequently for a network with near-neighbor transitions, 
which is like having ``connectors in series", we get 
\beq 
\barline{w} \ \ \equiv \ \ \left( \frac{1}{N} \sum_x \frac{1}{w(x)} \right)^{-1}
\eeq
with similar definitions for $\barline{w}^{\beta}$ and $\barline{w}^{\epsilon}$.
It is now natural to define a dimensionless driving intensity $\epsilon^2$ 
and dimensionless coupling parameters $g_n$, such that the latter 
reflect the relative exposure of the bonds to the driving:  
\beq
\frac{\barline{w}^{\epsilon}}{\barline{w}^{\beta}} \ \equiv &  \epsilon^2,  
&\ \ \ \ \ \mbox{[definition of $\epsilon^2$]}
\\ \label{e1002}
\frac{w_{\widebar{n}}^{\epsilon}}{w_{\widebar{n}}^{\beta}} \ \equiv &  g_n \epsilon^2,  
&\ \ \ \ \ \mbox{[definition of $g_n$]}
\eeq
{\em Sparsity} means that the couplings to the driving source 
are log-wide distributed.   
In the numerics we have assumed log-box distribution of the $g_n$ over several decades.
Namely, we the values $\ln(g_n)$ were generated such that 
they form a uniform distribution in the range ${[g_{\min},g_{\max}]}$.
As for the bath: the $w_{\widebar{n}}^{\beta}$ were assumed identical. 
Accordingly note that ${\overline{1/g_n}=1}$, 
while ${1 \ll \overline{g_n} \ll \overline{g_n^2}}$. 
The variance is $\mbox{Var}(g_n)=\overline{g_n^2}-\overline{g_n}^2$.

%%%%%%%%%%%%%%%%%%%%%%%%%%%%%%%%%%%%%%%%%%%%%%%%%%%%%%%%%%%%%%%%%%%%%%%%%%%%%%%%%%%%%%%%%%
\vspace{5mm}
{\bf Estimating the SMF.-- }
In the presence of driving the SMF is non zero:
\beq
\mathcal{E}_{\circlearrowleft} 
\ \ = \ \ 
\ln \left[ \frac{\prod_x^{\circlearrowleft}w_x}{\prod_x^{\circlearrowright} w_x} \right]
\ \ \equiv \ \ \ln  \frac{ [\circlearrowleft]}{[\circlearrowright]}
\eeq
where $\prod^{\circlearrowleft}$ is the product of all the $N$ anticlockwise rates, 
and $\prod^{\circlearrowright}$ is similarly defined.  
If $T_B\gg\Delta$ we get  
\be{-1}
\mathcal{E}_{\circlearrowleft} 
&=& 
\ln \left[ \frac
{\prod_n (w^{\epsilon}_n+ w^{\beta}_{\overrightarrow{n}}) }  
{\prod_n (w^{\epsilon}_n+ w^{\beta}_{\overleftarrow{n}}) } 
\right]
\\ \nonumber
&\approx&
\sum_n \ln \left[ \frac
{w_{n}^{\epsilon} + \left(1-\frac{\Delta_n}{2T_B} \right) w_{\widebar{n}}^{\beta}} 
{w_{n}^{\epsilon} + \left(1+\frac{\Delta_n}{2T_B} \right) w_{\widebar{n}}^{\beta}} 
\right]
\\ \label{e11}
&\approx& -\sum_{n} \left[ \frac{1}{1+g_n\epsilon^2} \right] \frac{\Delta_n}{T_B}
\eeq

Recall that we have ${\sum_n\Delta_n=0}$. Additionally we define 
\beq
\Delta^{(0)} &\equiv&  \sum_{n} g_n  \Delta_n 
\ \sim \ \pm \Big[2N \, \mbox{Var}(g)\Big]^{1/2} \Delta
\\
\Delta^{(\infty)} &\equiv& \sum_{n} \frac{1}{g_n}  \Delta_n
\ \sim \ \pm \Big[2N \, \mbox{Var}(g^{-1})\Big]^{1/2} \Delta
\eeq
The RMS-based estimate of the sums follows from the observation 
that, say, $\Delta^{(0)}$ can be rearranged as ${\sum_n (g_{n+1}-g_n) E_n}$, 
which is a sum of~$N$ independent random variables.  
Consequently we get for the SMF the following approximation 
\be{131}
\mathcal{E}_{\circlearrowleft} \ \ \approx \ \ 
\frac{1}{T_B}
\left\{
\amatrix{
\Delta^{(0)}\epsilon^2, & \ \ \epsilon^{2} < 1/g_{\max} \cr 
-\Delta^{(\infty)}/\epsilon^2, & \ \ \epsilon^{2} > 1/g_{\min} \cr
\sim[\pm] \Delta^{(*)}, & \ \ \mbox{otherwise} 
}\right.
\eeq
where $\Delta^{(*)}\equiv N^{1/2}\Delta$, and the $\sim$ implies 
that the the result exhibit fluctuations as $\epsilon^2$ is varied.
Looking in \Fig{f1a}, at the plot of $|\mathcal{E}_{\circlearrowleft}|$,  
we notice that there are dips that indicate that the SMF changes sign. 
These sign reversals depend on the specific realization of $g_{n}$ and $\Delta_{n}$.

%%%%%%%%%%%%%%%%%%%%%%%%%%%%%%%%%%%%%%

%%%%%%%%%%%%%%%%%%%%%%%%%%%%%%%%%%%%%%

%%%%%%%%%%%%%%%%%%%%%%%%%%%%%%%%%%%%%%%%%%%%%%%%%%%%%%%%%%%%%%%%%%%%%%%%%%%%%%%%%%%%%%%%%%
%%%%%%%%%%%%%%%%%%%%%%%%%%%%%%%%%%%%%%%%%%%%%%%%%%%%%%%%%%%%%%%%%%%%%%%%%%%%%%%%%%%%%%%%%%
\vspace{5mm}
{\bf The NESS current.-- }
The rate equation for nearest neighbor transitions is 
\be{2}
\dot{p}_n \ = \ w_{\overleftarrow{n{+}1}} p_{n+1} + w_{\overrightarrow{n}} p_{n-1} - 2w_{\widebar{n}}p_n
\ee
This set of equations is redundant because of conservation of probability.
At steady state  $\dot{p}_n = 0$, and there is some current $I$ in the system.
So we can write the equivalent non-redundant set of $N{+}1$ equations 
\be{13}
w_{\overrightarrow{n}}p_{n{-}1} - w_{\overleftarrow{n}}p_{n} = I, 
\ \ \ \ \ \ \ \sum_n p_n = 1
\eeq
This set of equations can be solved for the current using 
elementary algebra, or alternatively using the network formalism for stochastic systems \cite{net1,net2,net3}. 
The result may be written in compact notation as follows:
\be{7}
I \ \ &=& \ \ \frac{\prod_x^{\circlearrowleft}w_x - \prod_x^{\circlearrowright} w_x}{\sum_{x,n} \prod_{x'}^{(x \leadsto n)}  w_{x'}}
\ \ \equiv \ \ \frac{ [\circlearrowleft]-[\circlearrowright]}{[\leadsto]}
\ee
where the denominator has a sum over bonds~$x$ and target sites~$n$. 
In each term the bond~$x$ is disconnected to open the ring. 
The product $\prod^{(x \leadsto n)}$ denotes the multiplication of 
the ${N{-}1}$ rates that are directed from the cut point towards the target site $n$.
Note that the $N{-}1$ rates belong to the two paths that lead from~$x$ to~$n$.
In continuum limit notations the expression can be written as 
\beq
I \ = \ \frac{\eexp{\mathcal{E}_{\circlearrowleft}/2}-\eexp{-\mathcal{E}_{\circlearrowleft}/2} }
{\sum_{x} \frac{1}{w(x)} \int dx'\eexp{\mathcal{E}(x \leadsto x')/2}} 
\ee
Which can be roughly estimated as  
\be{54}
I \ \ \sim \ \ \frac{1}{N}\barline{w} 
\ \exp\left[-\frac{\mathcal{E}_{\cap}}{2}\right] 
\ 2\sinh\left(\frac{\mathcal{E}_{\circlearrowleft}}{2}\right)
\ee
This rough estimate is tested in \Fig{f1a}, and is in fact  quite satisfactory.

\begin{figure}

\includegraphics[width=\hsize]{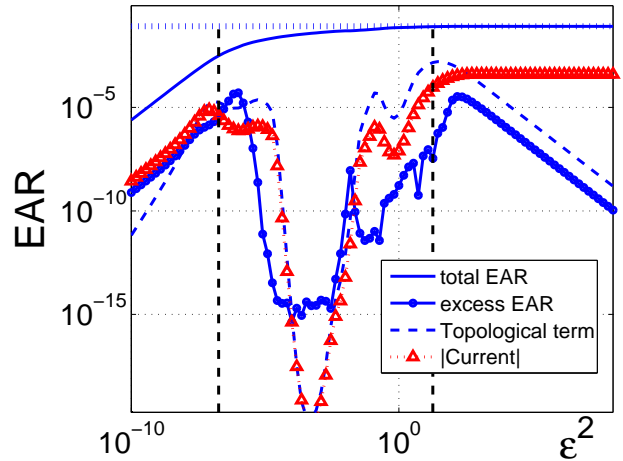}

\caption{The energy absorption rate (EAR) versus the driving intensity 
for a Ring with $N=10^{6}$ sites. The other parameters are 
the same as in the previous figure. The thin solid curve is the total EAR, 
while the dotted horizontal line is the expected saturation value.
The decorated solid curve is the EAR difference if the Ring is 
disconnected at one point, while the dashed curve is the topological 
term $\mathcal{E}_{ \circlearrowleft} I$.
The current in the ring is also drawn (triangular markers).
There is clearly a correlation between the excess EAR and the current.  
}
\label{f2}
\end{figure}

Let us discuss in more detail the current for very weak and very strong driving. 
In both limits the SMF becomes very small, 
the Sinai factor $\exp[]$ becomes of order unity,  
and the $\sinh()$ can be approximated by a linear function. 
Accordingly we get 
\be{55}
I \ \ \approx \ \ \frac{1}{N}
\left\{
\amatrix{
-[\Delta^{(0)}/T_B] \barline{w}^{\epsilon}, & \ \ \mbox{linear regime}
\cr
[\Delta^{(\infty)}/T_B] \barline{w}^{\beta}, & \ \ \mbox{Saturation} 
}\right.
\ee
As evident from these expressions, and as implied by the numerics, 
the direction of the current can change as the strength of the driving 
is increased, hence the dips in $|I|$ in \Fig{f2}. 
The small $\epsilon$ result is independent of $\barline{w}^{\beta}$, 
in spite of the bath dominance. The fingerprints 
of the bath show up only for strong driving, 
where the result becomes saturated, 
independent of $\barline{w}^{\epsilon}$.
The saturation by itself could have been anticipated: 
it is due to having SMF $\propto 1/\epsilon^2$ with rates $\propto \epsilon^2$.

In the intermediate regime $|\mathcal{E}_{\circlearrowleft}| \sim N^{1/2}$.
The maximal potential variation $\mathcal{E}_{\cap}$ is of the same order 
of magnitude, but always lager, typically by some factor of order unity. 
Hence the current becomes exponentially small in $\sqrt{N}$, 
as in the model by Sinai. Sparsity is the crucial requirement 
in order to observe this intermediate Sinai regime, 
otherwise there is merely a crossover from the linear response regime 
to the saturation regime.  

%%%%%%%%%%%%%%%%%%%%%%%%%%%%%%%%%%%%%%%%%%%%%%%%%%%%%%%%%%%%%%%%%%%%%%%%%%%%%%%%%%%%%%%%%%
%%%%%%%%%%%%%%%%%%%%%%%%%%%%%%%%%%%%%%%%%%%%%%%%%%%%%%%%%%%%%%%%%%%%%%%%%%%%%%%%%%%%%%%%%%
\vspace{5mm}
{\bf The EAR formula.-- }
By construction it is obvious that 
for zero driving intensity the system is detailed balanced 
with the temperature that is dictated by the bath: 
hence \Eq{e71} is satisfied with ${T_{nm}=T_B}$. 
Also for non zero driving the system 
is detailed balanced, {\em provided} the current is zero.    
This is implied by \Eq{e13}. Consequently, 
in the absence of current, \Eq{e71} is still 
satisfied, but the microscopic temperature of the $n$th 
transition is 
\beq
\frac{1}{T_n^{(0)}} \ = \ 
\frac{1}{\Delta_n}
\ln\left[\frac{w_{\overleftarrow{n}}}{w_{\overrightarrow{n}}}\right]
\eeq
For ${T_B \gg \Delta}$ one easily obtains the following practical approximation:
\beq
\frac{1}{T_n^{(0)}} \ \approx \ \left[\frac{w_{\overleftarrow{n}}-w_{\overrightarrow{n}}}{w_{\widebar{n}}}\right]\frac{1}{\Delta_n}
\ \approx \  \left[\frac{1}{1+g_n \epsilon^2} \right]\frac{1}{T_B}
\eeq
One observes that in the absence of current the system is  
locally heated, with microscopic temperatures ${T_n^{(0)}>T_B}$  
that are non-uniform due to the dispersion in the couplings. 

Let us see what happens if the current is non zero.
Again we can try to use \Eq{e71} in order to define 
a set of microscopic temperatures~$T_n$. 
One should notice that unlike~$T_n^{(0)}$, 
some $T_n$ might be smaller than ${T_B}$ or even 
negative reflecting a probability occupation inversion 
(in the sense of Laser physics).
For the following analysis it is useful to observe 
that \Eq{e13} provides an explicit expression for the 
occupation difference:  
\be{22}
p_{n{-}1}-p_{n} = 
\left[\frac{w_{\overleftarrow{n}}{-}w_{\overrightarrow{n}}}{w_{\overleftarrow{n}}{+}w_{\overrightarrow{n}}}\right]2\bar{p}_n  
+ \left[\frac{2}{w_{\overleftarrow{n}}{+}w_{\overrightarrow{n}}}\right] I
\eeq
where ${\bar{p}_n=(p_{n{-}1}+p_{n})/2}$ is defined 
as the average occupation of the sites in the $n$th bond. 
Note also that ${w_{\overleftarrow{n}}{-}w_{\overrightarrow{n}}}$ 
originates exclusively from the bath, because the 
driving induces symmetric transitions.

We now can proceed to derive an expression for the EAR, 
which equals to the rate in which energy goes from the 
system to the bath. 
Accordingly we have to calculate the 
rate of energy flow $\dot{\mathsf{Q}}$ that is associated with the bath 
induced transitions:
\be{0}
\dot{\mathsf{Q}} = \sum_n \left[ w^{\beta}_{\overleftarrow{n}} p_n - w^{\beta}_{\overrightarrow{n}} p_{n-1} \right]\Delta_n \eeq
Note that in steady state this should equal the rate of work $\dot{\mathsf{W}}$, 
that is associated with the driving source induced transitions. 
We would like now to simplify the above expression.
The first step is  to rewrite ${a_1b_1-a_2b_2}$
as the combination of $(a_1{-}a_2)(b_1{+}b_2)$ and $(a_1{+}a_2)(b_1{-}b_2)$.
This gives
\be{0}
\dot{\mathsf{Q}} = \sum_n \left[ (w_{\overleftarrow{n}}{-}w_{\overrightarrow{n}}) \bar{p}_{n} + w_{\widebar{n}}^{\beta} (p_{n}{-}p_{n{-}1})\right] \Delta_n
\eeq
For the calculation of the differences in the above expression
we use \Eq{e111} and \Eq{e22} respectively,
and further simplify by assuming $T_B\gg\Delta$, 
leading to 
\be{27}
\dot{\mathsf{Q}} =
\sum_n \left[\frac{\bar{p}_n w_{\widebar{n}}^{\beta}\Delta_n^2}{T_B}
-\frac{\bar{p}_n w_{\widebar{n}}^{\beta}\Delta_n^2}{T_n^{(0)}}
- I \frac{w_{\widebar{n}}^{\beta}}{w_{\widebar{n}}}\Delta_n\right] 
\eeq
In the above expression it was convenient to keep 
using the notation $T_n^{(0)}$ even if the current is non-zero.

The last term in the EAR expression \Eq{e27} arises 
because we have a non-trivial topology that can 
support non-zero current. We therefore name it `topological term'.  
We see that this term depends on the 
ratio ${w_{\widebar{n}}^{\beta}/(w_{\widebar{n}}^{\beta}+w_{\widebar{n}}^{\epsilon})}$.
Using the definition \Eq{e1002} and the expression for the SMF \Eq{e11} 
it takes the form ${T_B \mathcal{E}_{\circlearrowleft} I}$.
Consequently we can write the expression for the EAR 
schematically as in Ref.\cite{kbb} with 
an the additional {\em topological term}: 
\be{28}
\dot{\mathsf{Q}} 
\ \approx \ 
\left[ \frac{D_B}{T_B}- \frac{D_B}{T^{(0)}} \right]   
+ T_B\mathcal{E}_{\circlearrowleft} \ I
\eeq 
The first term represents the heat flow due to 
a temperature gradient. The definitions of the diffusion 
coefficient~$D_B$ and the induced temperature $T^{(0)}$ 
are implied by a comparison with \Eq{e27}.

It is interesting to point out that \Eq{e28} connects between the entropy
production obtained under different levels of coarse graining~\cite{eprd}. 
The term $\mathcal{E}_{\circlearrowleft} I$ is a coarse-grained entropy production 
for a setup in which one can not distinguish between transitions mediated by the
thermal bath and noise. In contrast, $\dot{Q}/T_B$ is the entropy production
rate when the two types of transitions are distinguishable.

%%%%%%%%%%%%%%%%%%%%%%%%%%%%%%%%%%%%%%%

%%%%%%%%%%%%%%%%%%%%%%%%%%%%%%%%%%%%%%%%%%%%%%%%%%%%%%%%%%%%%%%%%%%%%%%%%%%%%%%%%%%%%%%%%%
%%%%%%%%%%%%%%%%%%%%%%%%%%%%%%%%%%%%%%%%%%%%%%%%%%%%%%%%%%%%%%%%%%%%%%%%%%%%%%%%%%%%%%%%%%
\vspace{5mm}
{\bf EAR vs Current.-- }
We would like to inquire whether the EAR is correlated 
with the current. In a superficial glance one may be tempted 
to say that there is a linear correlation due to the 
topological term $\propto I$ in \Eq{e28}.
However, one should realize that also the $p_n$ in \Eq{e27}  
depend implicitly on the current. 
Hence one should not expect a strict linear relation,  
but, speculatively, a weaker correlation.  
To confirm this conjecture we calculate in \Fig{f2} 
the {\em difference} between the EAR of a connected Ring, 
and the EAR of the same ring after it had been disconnected 
at one point. We indeed observe that there is an unambiguous correlation. 

The driving induces non-vanishing current~$I$, 
hence the EAR of a closed ring is larger than  
that of a linear chain. In the linear 
response regime one may expand the EAR in powers 
of~$\epsilon^2$. One should observe that the $I$ related 
corrections are of order $\epsilon^4$. 
Let us write the explicit expression:
\beq
\dot{\mathsf{Q}} 
&=&  
\sum_n  \frac{\bar{p}_n w_{\widebar{n}}^{\beta}\Delta_n^2}{T_B}     
\left[1-\frac{1}{1+g_n \epsilon^2}\right]
+  T_B\mathcal{E}_{\circlearrowleft} \ I
\\
&\approx&  
\frac{D_B}{T_B}  
\left[ 
\overline{(g_n \epsilon^2) - (g_n \epsilon^2)^2} + \mbox{Var}(g) \epsilon^4\right] 
\label{eVarg}
\eeq 
When comparing a closed ring to a disconnected 
ring, one should be aware that the last term, 
which we call `topological term', should be either 
included or excluded respectively.
Compared with a disconnected ring the topological 
term obviously {\em increases} the EAR, but from 
the expression above we see that the global dependence 
of the EAR on the driving intensity remains sub-linear. 
Namely, one realizes that without the topological term  
the $\epsilon^4$ component in the EAR expression 
has the coefficient~$-\overline{g_n^2}$, 
while with the topological term the net coefficient 
becomes~$-\overline{g_n}^2$, 
which implies larger EAR but still sub-linear.

%%%%%%%%%%%%%%%%%%%%%%%%%%%%%%%%%%%%%%%%%%%%%%%%%%%%%%%%%%%%%%%%%%%%%%% 
%%%%%%%%%%%%%%%%%%%%%%%%%%%%%%%%%%%%%%%%%%%%%%%%%%%%%%%%%%%%%%%%%%%%%%%
\vspace{5mm}
{\bf Summary.-- }
The study of transport in network systems  
has numerous applications, notably in physical chemistry, 
where the dynamics is commonly described by a rate equation. 
There is much interest in studying NESS currents that are induced 
either by periodically varying system parameters \cite{saar}, 
or by {\em stochastic driving}.  
Assuming the latter, we have considered the NESS of a driven ring 
that is coupled to a bath, and found both the steady state 
current and the EAR.   
We have demonstrated how the ring-like topology 
and the sparsity lead to a glassy NESS with a non-trivial 
current dependence, 
exhibiting an interesting crossover from linear-response 
to saturation via an intermediate Sinai-type regime.

%%%%%%%%%%%%%%%%%%%%%%%%%%%%%%%%%%%%%%%%%%%
%%%%%%%%%%%%%%%%%%%%%%%%%%%%%%%%%%%%%%%%%%%
\vspace{5mm}
{\bf Acknowledgments.-- } 
This research was supported by the Israel Science Foundation (grant No.29/11).
DC~thanks Bernard Derrida for an insightful discussion.

%%%%%%%%%%%%%%%%%%%%%%%%%%%%%%%%%%%%%%%%%%%%%%%%%%%%%%%%%%%%%%%%%%%%%%%%%%%%%%%%%%%%%%%%%%%%%%%%%%%%%%%%
%%%%%%%%%%%%%%%%%%%%%%%%%%%%%%%%%%%%%%%%%%%%%%%%%%%%%%%%%%%%%%%%%%%%%%%%%%%%%%%%%%%%%%%%%%%%%%%%%%%%%%%%

%%%%%%%%%%%%%%%%%%%%%%%%%%%%%%%%%%%%%%
%%%%%%%%%%%%%%%%%%%%%%%%%%%%%%%%%%%%%%

%%%%%%%%%%%%%%%%%%%%%%%%%%%%%%%%%%%%%%%%%%%%%%%%%%%%%%%%%%%%%%%%%
%%%%%%%%%%%%%%%%%%%%%%%%%%%%%%%%%%%%%%%%%%%%%%%%%%%%%%%%%%%%%%%%%
\clearpage
\end{document}